\documentclass[12pt]{article}
\usepackage{times,amsmath,epsfig}
\usepackage{latexsym,amssymb}
\usepackage{cite}
\newcommand{\lettrineo}[2]{%
\settowidth{\lwidth}{#2\kern2pt}%
\noindent\hangindent\lwidth\hangafter-#1\hskip-\lwidth%
\smash{\hbox to\lwidth{\raise7pt\vtop{\null\hbox{#2}}%
\hfill}}\ignorespaces}

\newfont{\HUGEfonto}{cmr17 scaled \magstep3}  

\DeclareSymbolFont{AMSa}{U}{msa}{m}{n}
\DeclareMathSymbol{\blacksquare}  {\mathord}{AMSa}{"04}

\newfont{\bbb}{msbm10 scaled 500}

\newfont{\bb}{msbm10 scaled 1100}

\newcommand{\lambdav}{\hbox{\boldmath$\lambda$}}

\newcommand{\muv}{\hbox{\boldmath$\mu$}}

\renewcommand{\det}{{\hbox{det}}}

\newcommand{\beq}{\begin{equation}}
\newcommand{\enq}{\end{equation}}
\newcommand{\beqa}{\begin{eqnarray}}
\newcommand{\enqa}{\end{eqnarray}}

\newcommand{\beql}[1]{\begin{equation}\label{#1}}

\newcommand{\be}{\beta}

\newcommand{\qed}{\hfill $\Box$}

\newtheorem{thm}{Theorem}

\newtheorem{cor}{Corollary}

\newtheorem{defn}{Definition}

\newenvironment{proof}{{\sl Proof\/}:\ \ }{\qed\vspace{\baselineskip}}

\def\bbC{{\sf C}\kern -6pt {\sf C}}
\def\bbF{{\sf F}\kern -5pt {\sf F}}
\def\bbR{{\sf R}\kern -6pt {\sf R}}
\def\bbZ{{\sf Z}\kern -5pt {\sf Z}}
\def\sfbegin{\begingroup\sf}
\def\sfend{\endgroup}

\def\be{\begin{eqnarray*}}
\def\ee{\end{eqnarray*}}

\title{The Water-Filling Game in Fading Multiple Access Channels\footnote{The authors are with the ECE department at
The Ohio State University (\{lail,helgamal\}@ece.osu.edu). This
work was supported in part by the National Science Foundation and
Nokia Research Labs.}}
\author{Lifeng~Lai~and~Hesham~El~Gamal}
\begin{document}
\maketitle

\begin{abstract}
We adopt a game theoretic approach for the design and analysis of
distributed resource allocation algorithms in fading multiple
access channels. The users are assumed to be selfish, rational,
and limited by average power constraints. We show that the
sum-rate optimal point on the boundary of the multiple-access
channel capacity region is the unique Nash Equilibrium of the
corresponding water-filling game. This result sheds a new light on
the opportunistic communication principle and argues for the
fairness of the sum-rate optimal point, at least from a game
theoretic perspective. The base-station is then introduced as a
player interested in maximizing a weighted sum of the individual
rates. We propose a Stackelberg formulation in which the
base-station is the designated game leader. In this set-up, the
base-station announces first its strategy defined as the decoding
order of the different users, in the successive cancellation
receiver, as a function of the channel state. In the second stage,
the users compete conditioned on this particular decoding
strategy. We show that this formulation allows for achieving all
the corner points of the capacity region, in addition to the
sum-rate optimal point. On the negative side, we prove the
non-existence of a base-station strategy in this formulation that
achieves the rest of the boundary points. To overcome this
limitation, we present a repeated game approach which achieves the
capacity region of the fading multiple access channel. Finally, we
extend our study to vector channels highlighting interesting
differences between this scenario and the scalar channel case.
\end{abstract}
\section{Introduction}
The design and analysis of efficient resource allocation
algorithms for wireless channels has received significant research
interest for many years. In a pioneering work, Tse and Hanly have
characterized the capacity region of the fading multiple access
channel and the corresponding optimal power and rate allocation
policies \cite{Tse}. The centralized nature of these policies
motivates our work here on the design and analysis of distributed
allocation strategies that approach the optimal performance.
Arguably, such distributed implementations are more desirable from
a practical perspective.

In this paper, we adopt a game theoretic framework where the users
are typically modelled as rational and selfish players interested
in maximizing the utilities they obtain from the network. The
selfish behavior implies that individual users do not care about
the overall system performance. Over the last ten years, game
theoretic tools have been used to design distributed resource
allocation strategies in a variety of contexts. For example,
Mackenzie {\em et al.} consider the collision channel
\cite{Mackenzie}, Yu {\em et al.} focus on the digital subscriber
line setup \cite{weiyu}, Etkin {\em et al.} investigate the power
allocation game in the Gaussian interference channel \cite{Raul},
and La {\em et al.} model the power control problem in Gaussian
multiple access channels as a cooperative game where the users are
allowed to form coalitions \cite{Richard}. Probably the scenario
closest to our work is the design of distributed power control
algorithms for the up-link of Code Division Multiple Access (CDMA)
systems considered in e.g.,
\cite{CemNa,Farhad,Xiao,Zhou,Alpcan,Sung}. These papers focus on
{\bf time-invariant} channels and construct utility functions that
allow the users to reach a socially optimal equilibrium. These
works, however, reach the \textbf{negative} conclusion that the
selfish behavior entails a fundamental performance loss in the
sense that the achievable utilities at the equilibria
points\footnote{The rigorous definition of equilibria points will
be given in the sequel.}, if they exist, are usually inefficient
as compared with the centralized policy~\cite{CemNa,Alpcan}. The
central contribution of this paper is showing how to overcome this
negative conclusion in fading channels by exploiting the time
varying nature of fading, modelling the base-station as an
additional player with the appropriate decoding strategy, and
resorting to a repeated game formulation if needed.

We start with a static Nash formulation which only models the
multiple access users as players. In this formulation, every
player treats the signals of other users as Gaussian noise (with
the appropriate variance) and is interested in maximizing its
achievable rate subject to an average power constraint. The static
nature of the game implies that the game is played only once, and
{\bf not} a fixed channel environments. In this scenario, the
optimal power allocation strategy of every player is given by the
water-filling response to other players' strategies. Remarkably,
we show that the unique Nash equilibrium of this water-filling
game is the sum-rate optimal point on the boundary of the capacity
region \cite{Tse}. In a sense, this result establishes the
fairness of the sum-rate point, at least from a game theoretic
perspective. Hoping to achieve other boundary points of the
capacity region, we then introduce the base-station as a player
interested in maximizing a weighted sum of the individual rates.
By allowing the base-station to announce its decoding strategy
first, we transform our game into a Stackelberg formulation
\cite{mitgame}. Here, we establish the ability of this approach to
achieve all the corner points of the capacity region in addition
to the sum-rate optimal point. The key idea is for the
base-station to use a successive decoding strategy while altering
the decoding order as a function of the channel state. The final
step, that allows for achieving all points on the boundary of the
capacity region, is to use a dynamic game approach. In this
set-up, the base-station can use the decoding order as a {\em
punishment} tool forcing the multiple access users to adopt the
optimal power control policies. We then extend our results to
vector channels where different conclusions (as compared to the
scalar case) are drawn. It is worth noting that our approach is
purely information theoretic, and hence, we do not introduce other
elements such as pricing mechanisms~\cite{CemNa} into the problem.
In particular, we limit the payoff functions to depend only on the
achievable rate(s), and define the multiple access user strategy
as a power/rate allocation policy and the base-station strategy as
a decoding algorithm.

The rest of the paper is organized as follows. In
Section~\ref{back}, we present the system model and review,
briefly, known results on the capacity of fading multiple access
channels. Section~\ref{water} includes our results on the
water-filling game for scalar fading channels. In particular, we
devote Section~\ref{sec:basenp} to the Nash formulation,
Section~\ref{stack} to the Stackelberg formulation, and
Section~\ref{repeated} to the dynamic game scenario.
Section~\ref{vector} highlights some interesting structural
differences between scalar and vector channels. Finally, we close
with some concluding remarks in Section~\ref{conclusion}.

\section{Background} \label{back}
We consider a discrete-time flat fading multiple access channel
with $N$ users and one base-station. The signal received by the
base-station at time $n$ is\footnote{In this paper, we use lower
case letters for scalars, bold face lower case letters for vectors
and bold face upper case letters for matrices.}
\begin{equation}
y(n)=\sum \limits_{i=1}^{N} \sqrt{h_{i}(n)} x_{i}(n)+z(n),
\end{equation}
where $x_{i}(n)$ and $h_{i}(n)$ are the transmitted signal and
fading channel gain of the $i$th user at time $n$. Similar
to~\cite{Tse}, we assume the fading process to be jointly
stationary and ergodic. We further assume that the stationary
distribution has a continuous density and is bounded. User $i$ has
an average power constraint $\bar{P}_{i}$ and $z(n)$ is a sample
of a zero-mean white Gaussian noise process with variance
$\sigma^2$. The capacity region of this channel depends on the
fading process characteristics and the availability of the channel
state information (CSI).

\begin{figure}[thb]
\centering
\includegraphics[width=0.5 \textwidth]{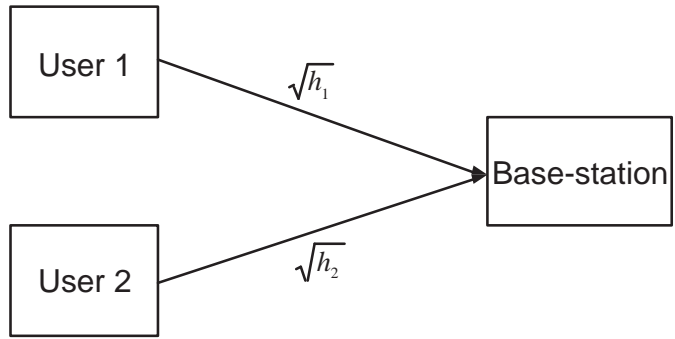}
\caption{The two-user multiple access channel.}
\label{fig:sysconfig}
\end{figure}

If the channel gains are assumed to be fixed and known {\em
a-priori} (i.e., time invariant channel) then we are reduced to
the Gaussian multiple-access channel where the capacity region is
well known~\cite{Cover}. For the two user case, this region
$\mathcal{G}_{g}$ is given by:
\begin{eqnarray}\label{equ:gauss}
R_{1}&\leq&\frac{1}{2}\log_{2}\bigg(1+\frac{h_{1}\bar{P}_{1}}{\sigma^2}\bigg),\nonumber\\
R_{2}&\leq&\frac{1}{2}\log_{2}\bigg(1+\frac{h_{2}\bar{P}_{2}}{\sigma^2}\bigg),\\
R_{1}+R_{2}&\leq&\frac{1}{2}\log_{2}\bigg(1+\frac{h_{1}\bar{P}_{1}+h_{2}\bar{P}_{2}}{\sigma^2}\bigg).\nonumber
\end{eqnarray}
It is easy to see that the boundary of $\mathcal{G}_{g}$ is a
pentagon. The two corner points are achieved by employing a
successive decoding strategy at the base-station and other
boundary points are achieved by appropriate time sharing between
the two decoding strategies used at the corner
points~\cite{Cover}. For time-varying channels with only receiver
CSI, the capacity region is also known~\cite{Gallager}. For the
two user case, the new capacity region can be interpreted as the
average of the rate expressions in~\eqref{equ:gauss} with respect
to the fading channel distribution.

In this paper, we consider time-varying channels where the CSI is
available {\em a-priori} at all the transmitters and the receiver.
This scenario was considered by Tse and Hanly~\cite{Tse} where
they characterized the capacity region $\mathcal{G}_{c}$ along
with the corresponding centralized power and rate allocation
policies ($\mathcal{P}_{c}$, $\mathcal{R}_{c}$). It was also shown
in \cite{Tse} that the power and rate allocation policies are
unique and each boundary point corresponds to the maximization of
a weighted sum of the individual rates. All the boundary points
are achieved by successive decoding, where the decoding order is
determined by the rate award vector $\muv$~\cite{Tse}.

\begin{figure}[thb]
\centering
\includegraphics[width=0.4 \textwidth]{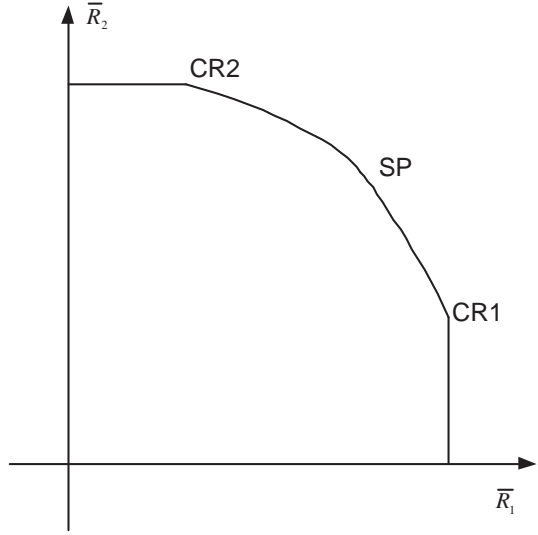}
\caption{The capacity region of the two user fading multiple
access channel.} \label{fig:tseregion}
\end{figure}

The capacity region for the two user case is shown in
Figure~\ref{fig:tseregion}. The corner point $CR_{1}$ is achieved
by using the following policy: user $1$ water-fills over the
background noise level and user $2$ water-fills over the sum of
the interference from user $1$ and the background noise. At the
base-station user $2$ is decoded first followed by user $1$. We
denote the rate pair at this point as
$(\bar{R}_{1,CR_1},\bar{R}_{2,CR_1})$. At point $CR_{2}$, the
roles of users $1$ and $2$ are reversed and we refer to the rate
pair by $(\bar{R}_{1,CR_2},\bar{R}_{2,CR_2})$. Another boundary
point of particular interest is the maximum sum-rate point $SP$.
Unlike the AWGN Multiple Access Channel (MAC), this point is
unique in our case and is achieved by a time-sharing policy where
only one user is allowed to transmit at any fading
state~\cite{Tse,Knopp}. This observation will prove instrumental
to the development of the main result in Section~\ref{sec:basenp}.

The centralized nature of the optimal power and rate allocation
policies ($\mathcal{P}_{c}$, $\mathcal{R}_{c}$) motivates our
pursuit for distributed strategies that approach the capacity
region of the fading MAC. Our assumption that the CSI is known
everywhere implies that the games considered here are games with
perfect and complete information~\cite{Mackenzie, weiyu,Raul,
Richard,CemNa,Farhad,Xiao,Zhou,Alpcan,Sung}. Without loss of
generality, and to avoid some tedious details, we limit our
discussion to pure strategies~\cite{Basar,mitgame}.

\section{The Water-Filling Game}\label{water}
For simplicity of presentation, we first consider in details the
two user scenario. Our arguments extend to the $N$ user channel as
briefly outlined in Section~\ref{arbitrary}.
\subsection{Nash Formulation}\label{sec:basenp}
Here, we consider a static non-cooperative game where the players
are the multiple-access users. In this game, the strategy of user
$i$ is the power control policy $\mathcal{P}_{i}$ and rate control
policy $\mathcal{R}_{i}$. The corresponding payoff function is
defined as the average achievable rate
$\bar{R}_{i}=E_{\mathbf{h}}[\mathcal{R}_{i}]$ with
$\mathbf{h}=[h_{1},h_{2}]^T$. The goal of user $i$ is to
\begin{eqnarray}
\max_{\mathcal{P}_{i}}
\bar{R}_{i}(\mathcal{P}_{i},\mathcal{P}_{-i}) \text{ s.t. }
\mathcal{P}_{i}\in \mathcal{F}_{i},
\end{eqnarray}
where
$\mathcal{F}_{i}=\{\mathcal{P}_{i}:E_{\mathbf{h}}[\mathcal{P}_{i}]\leq\bar{P}_{i},
\mathcal{P}_{i}(\mathbf{h})\geq 0\}$ is the set of all feasible
power control policies of user $i$, and $\mathcal{P}_{-i}$
represents the power control policy of the other user (in the more
general $\mathcal{P}_{-i}$ refers to the strategies of all users
except user $i$). Since the base-station is not a player of the
game, we assume that each user will treat the signal of the other
user as interference. Given the power control policy
$\mathcal{P}_{2}(h_{1},h_{2})$ of user 2, the payoff of user 1 is
given by
\begin{eqnarray}
\bar{R}_{1}&=&\int\int
\frac{1}{2}\log_{2}\Big(1+\frac{\mathcal{P}_{1}(h_{1},h_{2})h_{1}}{\sigma^2+\mathcal{P}_{2}(h_{1},h_{2})h_{2}}\Big)f(h_{1},h_{2})dh_{1}dh_{2}.
\end{eqnarray}
Here $f(h_{1},h_{2})$ is the joint probability density function of
the two fading coefficients. The payoff function of user 2 is
defined similarly. As we can see the payoff function of each user
depends on the two power control policies ($\mathcal{P}_{1}$,
$\mathcal{P}_{2}$). Before proceeding further, we need the
following definition from~\cite{Basar}.
\begin{defn}
A Nash equilibrium is a policy pair
$(\mathcal{P}_{1}^{*},\mathcal{P}_{2}^{*})$ such that
\begin{eqnarray}
\bar{R}_{1}(\mathcal{P}_{1}^{*},\mathcal{P}_{2}^{*})&\geq&
\bar{R}_{1}(\mathcal{P}_{1}^{'},\mathcal{P}_{2}^{*}),
\quad \forall \mathcal{P}_{1}^{'}\in \mathcal{F}_{1},\nonumber\\
\bar{R}_{2}(\mathcal{P}_{1}^{*},\mathcal{P}_{2}^{*})&\geq&
\bar{R}_{2}(\mathcal{P}_{1}^{*},\mathcal{P}_{2}^{'}), \quad\forall
\mathcal{P}_{2}^{'}\in \mathcal{F}_{2}.
\end{eqnarray}
\end{defn}
This definition means that at the Nash equilibrium, no user can
benefit by unilaterally deviating. Given a fixed power control
policy of user $2$, the optimal strategy
$\mathcal{P}_{1}(h_{1},h_{2})$ of user $1$ is the solution to the
following optimization problem
\begin{eqnarray}\label{opt}
\bar{R}_{1}&=&\max\limits_{\mathcal{P}_{1}} \int\int
\frac{1}{2}\log_{2}\Big(1+\frac{\mathcal{P}_{1}(h_{1},h_{2})h_{1}}{\sigma^2+\mathcal{P}_{2}(h_{1},h_{2})h_{2}}\Big)f(h_{1},h_{2})dh_{1}dh_{2},\nonumber\\
\text{s.t.}&&\int \int
\mathcal{P}_{1}(h_{1},h_{2})f(h_{1},h_{2})dh_{1}dh_{2}\leq
\bar{P}_{1},\\\nonumber &&\mathcal{P}_{1}(h_{1},h_{2})\geq 0.
\end{eqnarray}
We wish to emphasize the fact that each user is actually not aware
of the policy used by the other user. Starting from an arbitrary
initial point, each user can only rely on the assumption of
\emph{rationality} to \emph{guess} the policy employed by the
other user. Based on this guess, each user chooses a new policy as
a best response to the conceived policy of the other user. This
process is then repeated, hoping to converge at an {\em
equilibrium}. One of the central themes in game theory is to
characterize such equilibria, if they exist~\cite{mitgame}.

It is easy to verify that the objective function in (\ref{opt}) is
concave, the constraint set is convex, the Slater's condition is
satisfied, and hence, the solution to this problem is the
well-known water-filling power allocation, i.e.,
\begin{equation}
\mathcal{P}_{1}(h_{1},h_{2})=\Big(\lambda_{1}-\frac{\sigma^2}{h_{1}}-\frac{\mathcal{P}_{2}(h_{1},h_{2})h_{2}}{h_{1}}\Big)^+,
\end{equation}
in which $(x)^+=\max\{x,0\}$ and $\lambda_{1}$ is the power level
that satisfies
\begin{equation}
\int\int\Big(\lambda_{1}-\frac{\sigma^2}{h_{1}}-\frac{\mathcal{P}_{2}(h_{1},h_{2})h_{2}}{h_{1}}\Big)^+f(h_{1},h_{2})dh_{1}dh_{2}=\bar{P}_{1}.
\end{equation}

Similarly the optimal policy of user $2$, given a fixed policy for
user $1$, is given by
\begin{equation}
\mathcal{P}_{2}(h_{1},h_{2})=\Big(\lambda_{2}-\frac{\sigma^2}{h_{2}}-\frac{\mathcal{P}_{1}(h_{1},h_{2})h_{1}}{h_{2}}\Big)^+.
\end{equation}

From these expressions, one can see that the optimal policy of
each user depends largely on its {\em guess} of the other user
policy. Based on this guess, each user will determine its policy
and adjusts its water-filling level to maximize its own average
rate. At the Nash equilibrium, the water-filling pair
$(\lambda_{1},\lambda_{2})$ satisfies the two average power
constraints with equality. Now we are ready to prove our first
result.
\begin{thm}\label{thm:nashsum}
The maximum sum-rate point $SP$ of the capacity region
$\mathcal{G}_{c}$ is the unique Nash equilibrium of our
water-filling game.\end{thm}
\begin{proof}
At first, let's show the existence of only time-sharing
equilibria. Suppose there exists a non time sharing equilibrium
with the corresponding water-level pair
$(\lambda_{1},\lambda_{2})$. Then for some channel realizations
$h_{1}$, $h_{2}$, we have $\mathcal{P}_{1}(h_{1},h_{2})>0$,
$\mathcal{P}_{2}(h_{1},h_{2})>0$, and
\begin{equation}\label{equ:u1}
\frac{\sigma^2}{h_{1}}+\frac{\mathcal{P}_{2}(h_{1},h_{2})h_{2}}{h_{1}}+\mathcal{P}_{1}(h_{1},h_{2})=\lambda_{1},\nonumber
\end{equation}
\begin{equation}\label{equ:u2}
\frac{\sigma^2}{h_{2}}+\frac{\mathcal{P}_{1}(h_{1},h_{2})h_{1}}{h_{2}}+\mathcal{P}_{2}(h_{1},h_{2})=\lambda_{2}.
\end{equation}

From these two equations, we get
\begin{equation}\label{equ:u3}
\lambda_{1}=\lambda_{2}\frac{h_{2}}{h_{1}}.
\end{equation}

Since $\lambda_{1},\lambda_{2}$ are constants, and the fading
coefficients are characterized by a continuous pdf,~\eqref{equ:u3}
is satisfied with a zero probability. This implies the existence
of only time-sharing Nash equilibria.

Under the time-sharing equilibrium, when
$\mathcal{P}_{1}(h_{1},h_{2})>0$, the sum of the background noise
and the interference from user $1$ should be larger than the
water-level of user $2$. Thus when user $1$ transmits, the channel
conditions should satisfy the following inequality
\begin{equation}
\mathcal{P}_{1}(h_{1},h_{2})\frac{h_{1}}{h_{2}}+\frac{\sigma^{2}}{h_{2}}
=\Big(\lambda_{1}-\frac{\sigma^2}{h_{1}}\Big)\frac{h_{1}}{h_{2}}+\frac{\sigma^{2}}{h_{2}}=\frac{\lambda_{1}h_{1}}{h_{2}}\geq
\lambda_{2}.
\end{equation}

Similarly, when user $2$ transmits, the channel conditions should
satisfy the following condition
\begin{equation}
\frac{\lambda_{2}h_{2}}{h_{1}}\geq \lambda_{1}.
\end{equation}

The water-filling levels can now be obtained by solving the
following two equations
\begin{eqnarray}\label{equ:waterlevel}
\int_{0}^{\infty}\int_{\frac{\lambda_{2}h_{2}}{\lambda_{1}}}^{\infty}
\bigg(\lambda_{1}-\frac{\sigma^2}{h_{1}}\bigg)^{+}f(h_{1},h_{2})dh_{1}dh_{2}&=&\bar{P}_{1},\nonumber\\
\int_{0}^{\infty}\int_{\frac{\lambda_{1}h_{1}}{\lambda_{2}}}^{\infty}
\bigg(\lambda_{2}-\frac{\sigma^2}{h_{2}}\bigg)^{+}f(h_{1},h_{2})dh_{1}dh_{2}&=&\bar{P}_{2}.
\end{eqnarray}

The corresponding power control policies are unique and given by
\begin{equation}\label{equ:p1}
\mathcal{P}_{1}(h_{1},h_{2})=\Big(\lambda_{1}-\frac{\sigma^2}{h_{1}}\Big)^+,\text{
when}\; h_{1}\geq\frac{h_{2}\lambda_{2}}{\lambda_{1}},
\end{equation}

\begin{equation}\label{equ:p2}
\mathcal{P}_{2}(h_{1},h_{2})=\Big(\lambda_{2}-\frac{\sigma^2}{h_{2}}\Big)^+,\text{
when}\; h_{2}\geq\frac{h_{1}\lambda_{1}}{\lambda_{2}},
\end{equation}
with $\mathcal{P}_{1}(h_{1},h_{2})=0$ and
$\mathcal{P}_{2}(h_{1},h_{2})=0$ in other cases.

It was shown in~\cite{Tse} that centralized policy corresponding
to the point $SP$ is time sharing with the same power allocation
levels as~\eqref{equ:p1}~\eqref{equ:p2}. Finally, the fact that
the solution to~\eqref{equ:waterlevel} is unique \cite{Tse}
implies that the only Nash equilibrium of the distributed power
control game is the maximum sum-rate point of the capacity region
(i.e., $SP$).
\end{proof}

Two comments are now in order. \begin{enumerate}

\item Theorem~\ref{thm:nashsum} establishes the remarkable fact
that the selfish behavior of the users will lead them to {\bf
jointly} optimize the sum-rate of the channel. In fact, this
result provides a new interpretation of the opportunistic
communication principle~\cite{Knopp}. At any particular instance,
the user with the strongest channel sees a relatively weak
interference from the other user, and hence, decides to transmit
with a high power level. On the other hard, the other user sees a
strong interferer in addition to a weak channel, and hence,
decides to conserve the power for later usage. This way, they
reach the {\em opportunistic} time sharing equilibrium
distributively. This result also establishes a certain
game-theoretic fairness of the point $SP$. The underlying idea is
that the selfishness of the different users will {\em balance-out}
at the sum-rate optimal point. To impose other fairness criteria,
the base-station must be involved in the game as argued in the
next section.

\item Theorem~\ref{thm:nashsum} contrasts the negative conclusions
drawn in earlier works on the efficiency of game theoretic
approaches in CDMA up-link power control (e.g.,
\cite{CemNa,Farhad,Xiao,Zhou,Alpcan,Sung}). The enabling vehicle
behind this result is the time varying nature of the fading
channel. With this temporal variations, the CSI (available at all
transmitter) acts like a common randomness that allows the users
to reach a more efficient equilibrium based on a selfish
rationale. This is yet another manifestation of the positive
impact that fading, if properly exploited, can have on certain
aspects of wireless systems.

\end{enumerate}
\subsection{Stackelberg Formulation}\label{stack}

In the previous section, we have shown that the only boundary
point achievable by our Nash game is the optimal sum-rate point.
One can attribute this limitation to the assumption that every
user (player) will treat the other user's signal as noise. While
this assumption does not entail a loss at the {\em time sharing}
point $SP$, it does not allow for achieving other boundary points.
Such points require the base-station to employ a more
sophisticated decoding rule. In \cite{Tse}, it was shown that
successive decoding, with the appropriate ordering, is sufficient
to achieve all the boundary points.  This observation motivates a
game theoretic formulation where the base-station is introduced as
an additional player. The base-station strategy corresponds to a
particular choice of the decoding order, as detailed next.


We wish to stress that, unlike the centralized scenario
\cite{Tse}, the base-station in our formulation does not dictate
the power level and rate of the individual users. Still, it is
reasonable to assume that the roles of the base-station and
multiple-access users are not totally symmetric. Therefore, we do
not model the base-station as an {\em ordinary} player in our game
but rather appeal to the bi-level programming
notion~\cite{bilevel}. Bi-level programming is typically used in
modelling a decision making process where there is a hierarchical
relationship between the decision makers. In our context, bi-level
programming corresponds to a Stackelberg
game~\cite{bilevel,Basar}, where the leader announces its strategy
first and then the remaining players react according to a specific
equilibrium concept among them. Here, we designate the
base-station as the game leader, and hence, it will announce its
decoding strategy in the first level of the game. This way, the
base-station can rely on the rational and selfish nature of the
multiple access players to {\em influence} their behavior in the
second stage (i.e., low level game).

In this work, we consider a class of successive decoding
strategies parameterized by the decoding order as a function of
the fading gains $(h_{1},h_{2})$. More precisely, the base-station
divides the whole possible space of $(h_{1},h_{2})$ into two
subsets $D_{1},D_{1}^{c}$. When $(h_{1},h_{2})\in D_{1}$, the
base-station will decode user $1$'s information first whereas
$(h_{1},h_{2})\in D_{1}^c$ implies decoding user $2$'s signal
first. After the base-station announces its strategy, i.e.,
$D_{1}$, the multiple access users play the low level game using
the Nash equilibrium concept. The strategy space of user $i$ is
still $\mathcal{F}_{i}$, and the payoff function of user $i$ is
defined as the supremum of the achievable rate. Here supremum
refers to the fact that in the rate expressions to follow we
always assume the users to be decoded successfully (which is a
critical assumption in the successive decoding approach). We will
show later that, at the Nash equilibrium this condition indeed
holds. Hence, the supremum corresponds exactly to the achieved
payoff. With a slight abuse of notation, the payoff function of
each user is written as
\begin{eqnarray}\label{equ:game}
\bar{R}_{1}(D_{1},\mathcal{P}_{1},\mathcal{P}_{2})=\int\int\frac{1}{2}\log_{2}
\bigg(1+\frac{\mathcal{P}_{1}(h_{1},h_{2})h_{1}}{\sigma^2+\mathcal{P}_{2}(h_{1},h_{2})h_{2}I_{\{(h_{1},h_{2})\in
D_{1}\}}}\bigg) f(h_{1},h_{2})d h_{1} d h_{2},\\\nonumber
\bar{R}_{2}(D_{1},\mathcal{P}_{1},\mathcal{P}_{2})=\int\int\frac{1}{2}\log_{2}
\bigg(1+\frac{\mathcal{P}_{2}(h_{1},h_{2})h_{2}}{\sigma^2+\mathcal{P}_{1}(h_{1},h_{2})h_{1}I_{\{(h_{1},h_{2})\in
D_{1}^{c}\}}}\bigg)f(h_{1},h_{2})d h_{1} d h_{2}.
\end{eqnarray}
Here $I_{\{\cdot\}}$ is the indication function. In order to
achieve the average rate in (\ref{equ:game}), for a given
base-station strategy $D_{1}$, each user will use two code-books.
The low rate codebook is multiplexed across the fading states in
which the user is decoded first and the high rate codebook is
multiplexed across the other fading states. The payoff function of
the base-station is defined as
\begin{equation}
\mu_{1}\bar{R}_{1}(D_{1},\mathcal{P}_{1},\mathcal{P}_{2})+\mu_{2}\bar{R}_{2}(D_{1},\mathcal{P}_{1},\mathcal{P}_{2}).
\end{equation}
This payoff function has a natural economical interpretation as
the revenue of the base-station where $\mu_{i}$ can be viewed as
the payment that user $i$ owes per unit rate. The value of
$\mu_{i}$ can be decided using an auction
process~\cite{Sun.auction}, where each user submits its proposed
payment $\mu_{i}$ to the base-station in order to maximize its own
utility. In this work, we do not consider this auction process and
assume that $\muv=[\mu_1,\mu_2]^T$ is given.

We first study the properties of the low level game. The Nash
equilibrium under a fixed base-station strategy $D_{1}$ is a power
control pair $(\mathcal{P}_{1}^{*},\mathcal{P}_{2}^{*})$ that
satisfies
\begin{eqnarray}
\bar{R}_{1}(D_{1},\mathcal{P}_{1}^{*},\mathcal{P}_{2}^{*})&\geq&
\bar{R}_{1}(D_{1},\mathcal{P}_{1}^{'},\mathcal{P}_{2}^{*}),\;
\forall \mathcal{P}_{1}^{'}\in \mathcal{F}_{1},\nonumber\\
\bar{R}_{2}(D_{1},\mathcal{P}_{1}^{*},\mathcal{P}_{2}^{*})&\geq&
\bar{R}_{2}(D_{1},\mathcal{P}_{1}^{*},\mathcal{P}_{2}^{'}),\;\forall
\mathcal{P}_{2}^{'}\in \mathcal{F}_{2}. \nonumber
\end{eqnarray}

For any given power control policy $\mathcal{P}_{2}$, the optimal
power control policy of user $1$ is the solution to the following
optimization problem
\begin{eqnarray}\label{equ:userone}
\nonumber \max\limits_{\mathcal{P}_{1}}\quad
\bar{R}_{1}(D_{1},\mathcal{P}_{1},\mathcal{P}_{2})&=&\int\int\frac{1}{2}\log_{2}
\bigg(1+\frac{\mathcal{P}_{1}(h_{1},h_{2})h_{1}}{\sigma^2+\mathcal{P}_{2}(h_{1},h_{2})h_{2}I_{\{(h_{1},h_{2})\in D_{1}\}}}\bigg) f(h_{1},h_{2})d h_{1} d h_{2},\\
s.t. &&\int \int
\mathcal{P}_{1}(h_{1},h_{2})f(h_{1},h_{2})dh_{1}dh_{2}\leq
\bar{P}_{1},\\\nonumber &&\mathcal{P}_{1}(h_{1},h_{2})\geq 0.
\end{eqnarray}
The optimal power control policy of user 2 is also the solution to
a similar optimization problem for any power control policy of
user 1. For a given $D_{1}$, the solution set for this low level
game is written as
$S(D_{1})=\{(\mathcal{P}_{1},\mathcal{P}_{2}):(\mathcal{P}_{1},\mathcal{P}_{2})\;
\text{is a Nash equilibrium of the low level game}\}$. The
following result characterizes the pure-strategy Nash equilibria
of our low level game. The algorithm developed in the proof is
reminiscent of the iterative algorithm in~\cite{Tse,weiyu}.
\begin{thm}\label{thm:nash}
For any strategy $D_{1}$ of the base-station, and any channel
distribution, there exist Nash equilibria for the low level
distributed power/rate control game.
\end{thm}
\begin{proof}
At the Nash equilibrium, no user can benefit by deviating
unilaterally. Suppose $\mathcal{P}_{2}(h_{1},h_{2})$ is given,
user $1$'s strategy is the solution to~\eqref{equ:userone}, which
is still the water-filling solution
\begin{eqnarray}\label{equ:power1}
\mathcal{P}_{1}(h_{1},h_{2})=\bigg(\lambda_{1}-\frac{\sigma^2}{h_{1}}-\frac{\mathcal{P}_{2}(h_{1},h_{2})h_{2}I_{\{(h_{1},h_{2})\in
D_{1}\}}}{h_{1}}\bigg)^+,
\end{eqnarray}
where $\lambda_{1}$ is the power level chosen to satisfy the power
constraint of user $1$ with equality. For the same reason, if we
fix $\mathcal{P}_{1}(h_{1},h_{2})$, the optimal response of user 2
is also water-filling over the sum of the interference from user 1
and the background noise, which is
\begin{eqnarray}\label{equ:power2}
\mathcal{P}_{2}(h_{1},h_{2})=\bigg(\lambda_{2}-\frac{\sigma^2}{h_{2}}-\frac{\mathcal{P}_{1}(h_{1},h_{2})h_{1}I_{\{(h_{1},h_{2})\in
D_{1}^c\}}}{h_{2}}\bigg)^+.
\end{eqnarray}

The key of our proof is to establish the existence of a pair
($\lambda_{1},\lambda_{2}$) that simultaneously satisfies the two
power constraints with equality, and hence, constitutes a Nash
equilibrium. If such $(\lambda_{1},\lambda_{2})$ exists, we have
solutions to the equations~\eqref{equ:power1}
and~\eqref{equ:power2}. One can easily check that if
$(h_{1},h_{2})\in D_{1}$,
\begin{eqnarray}\label{equ:power1sim}
\mathcal{P}_{2}(h_{1},h_{2})&=&\bigg(\lambda_{2}-\frac{\sigma^2}{h_{2}}\bigg)^{+},\nonumber\\
\mathcal{P}_{1}(h_{1},h_{2})&=&\bigg(\lambda_{1}-\frac{\sigma^2}{h_{1}}-\frac{\mathcal{P}_{2}(h_{1},h_{2})h_{2}}{h_{1}}\bigg)^+\\\nonumber
&=&\bigg(\lambda_{1}-\frac{\sigma^2}{h_{1}}-\bigg(\frac{\lambda_{2}h_{2}}{h_{1}}-\frac{\sigma^2}{h_{1}}\bigg)^+\bigg)^+.
\end{eqnarray}
Similarly, if $(h_{1},h_{2})\in D_{1}^{c}$,
\begin{eqnarray}\label{equ:u2first}
\mathcal{P}_{1}(h_{1},h_{2})&=&\bigg(\lambda_{1}-\frac{\sigma^2}{h_{1}}\bigg)^{+},\nonumber\\
\mathcal{P}_{2}(h_{1},h_{2})&=&\bigg(\lambda_{2}-\frac{\sigma^2}{h_{2}}-\frac{\mathcal{P}_{1}(h_{1},h_{2})h_{1}}{h_{2}}\bigg)^+\\
&=&\bigg(\lambda_{2}-\frac{\sigma^2}{h_{2}}-\bigg(\frac{\lambda_{1}h_{1}}{h_{2}}-\frac{\sigma^2}{h_{2}}\bigg)^+\bigg)^+.\nonumber
\end{eqnarray}

Thus, if the water-filling level pair $(\lambda_{1},\lambda_{2})$
exists, it should be the solution to the following equation array:
\begin{eqnarray}\label{equ:lambda}
\iint \limits_{D_{1}}\bigg(\lambda_{1}-\frac{\sigma^2}{h_{1}}
-\bigg(\frac{\lambda_{2}h_{2}}{h_{1}}-\frac{\sigma^2}{h_{1}}\bigg)^+\bigg)^+f(h_{1},h_{2})d
h_{1} d h_{2}
\nonumber\\+\iint\limits_{D_{1}^{c}}\bigg(\lambda_{1}-\frac{\sigma^2}{h_{1}}\bigg)^+f(h_{1},h_{2})dh_{1}dh_{2}=\bar{P}_{1},
\end{eqnarray}
\begin{eqnarray}
\iint\limits_{D_{1}^{c}}\bigg(\lambda_{2}-\frac{\sigma^2}{h_{2}}-
\bigg(\frac{\lambda_{1}h_{1}}{h_{2}}-\frac{\sigma^2}{h_{2}}\bigg)^+\bigg)^+f(h_{1},h_{2})dh_{1}dh_{2}\nonumber\\\nonumber
+\iint\limits_{D_{1}}\bigg(\lambda_{2}-\frac{\sigma^2}{h_{2}}\bigg)^{+}f(h_{1},h_{2})dh_{1}dh_{2}=\bar{P}_{2}.
\end{eqnarray}

Before proceeding further, we first observe the following. If
there are two pairs $(\lambda_{1}^{'},\lambda_{2}^{'})$ and
$(\lambda_{1},\lambda_{2})$, where $\lambda_{1}^{'}>\lambda_{1},
\lambda_{2}^{'}=\lambda_{2}$, then we have
$\bar{P}_{1}(\lambda_{1}^{'},\lambda_{2}^{'})\geq
\bar{P}_{1}(\lambda_{1},\lambda_{2}),\bar{P}_{2}(\lambda_{1}^{'},\lambda_{2}^{'})\leq
\bar{P}_{2}(\lambda_{1},\lambda_{2})$\footnote{Here
$\bar{P}_{i}(\lambda_{1},\lambda_{2})$ refers to the average power
of user $i$ when the users do water-filling according to the water
levels $(\lambda_{1},\lambda_{2})$.}. One can easily verify this
by observing that $\mathcal{P}_{1}(h_{1},h_{2})$ is a
non-decreasing function of $\lambda_{1}$ and a non-increasing
function of $\lambda_{2}$. At the same time,
$\mathcal{P}_{2}(h_{1},h_{2})$ is a non-increasing function of
$\lambda_{1}$ and a non-decreasing function of $\lambda_{2}$.
Based on these observations, we have the following iterative
method to solve~\eqref{equ:lambda}. Set
$\lambda_{1}(1)=0,\lambda_{2}(1)=0$, then fix $\lambda_{2}$ and
increase $\lambda_{1}$ until
$\bar{P}_{1}(\lambda_{1},\lambda_{2}(1))=\bar{P}_{1}$. This can be
done by solving the following equation:
\begin{eqnarray}
\iint \limits_{D_{1}}\bigg(\lambda_{1}-\frac{\sigma^2}{h_{1}}
-\bigg(\frac{\lambda_{2}(1)h_{2}}{h_{1}}-\frac{\sigma^2}{h_{1}}\bigg)^+\bigg)^+f(h_{1},h_{2})d
h_{1} d h_{2}
\nonumber\\+\iint\limits_{D_{1}^{c}}\bigg(\lambda_{1}-\frac{\sigma^2}{h_{1}}\bigg)^+f(h_{1},h_{2})dh_{1}dh_{2}=\bar{P}_{1}.
\end{eqnarray}
Let $\lambda_{1}(2)$ represent the solution to this equation. At
this time, we will have
$\bar{P}_{2}(\lambda_{1}(2),\lambda_{2}(1))\leq\bar{P}_{2}$. Then
we can increase $\lambda_{2}(1)$ to $\lambda_{2}(2)$ such that
$\bar{P}_{2}(\lambda_{1}(2),\lambda_{2}(2))=\bar{P}_{2}$. After
this step, $\bar{P}_{1}(\lambda_{1}(2),\lambda_{2}(2))\leq
\bar{P}_{1}$, thus we can increase $\lambda_{1}$ again. Through
this process, we can get non-decreasing sequences
$\lambda_{1}(n),\lambda_{2}(n)$, and
$\bar{P}_{1}(\lambda_{1}(n),\lambda_{2}(n))\rightarrow\bar{P}_{1},
\bar{P}_{2}(\lambda_{1}(n),\lambda_{2}(n))\rightarrow\bar{P}_{2}$.
Since $\bar{P}_{1}, \bar{P}_{2}$ are limited,
$\lambda_{1}(n),\lambda_{2}(n)$ are non-decreasing sequence with
upper bounds. Then there exists constants
$\lambda_{1}^{*},\lambda_{2}^{*}$ such that:
\begin{eqnarray}
\lim \limits_{n\rightarrow \infty}
\lambda_{1}(n)&=&\lambda_{1}^{*},\quad\quad
\bar{P}_{1}(\lambda_{1}^{*},\lambda_{2}^{*})=\bar{P}_{1}.\\
\lim \limits_{n\rightarrow \infty}
\lambda_{2}(n)&=&\lambda_{2}^{*},\quad\quad
\bar{P}_{2}(\lambda_{1}^{*},\lambda_{2}^{*})=\bar{P}_{2}.
\end{eqnarray}
This pair $(\lambda_{1}^{*},\lambda_{2}^{*})$ is therefore a Nash
equilibrium of our power allocation game.
\end{proof}

Theorem~\ref{thm:nash} only establishes the existence of a Nash
equilibrium, but it tells nothing about the uniqueness of this
equilibrium. To prove uniqueness, one is typically forced to find
a contraction mapping whose fixed point is the Nash equilibrium.
In \cite{weiyu,Raul}, the authors apply this method to the
interference game and find that uniqueness requires very
restrictive conditions. Fortunately, we are able to prove
uniqueness in our setup by using the concept of
\textbf{admissible} Nash equilibrium (Definition 3.3
of~\cite{Basar}).

\begin{defn}
A Nash equilibrium strategy pair
$(\mathcal{P}_{1}^*,\mathcal{P}_{2}^*)$ is said to be admissible
if there exists no other Nash equilibrium strategy pair
$(\mathcal{P}_{1}^{'},\mathcal{P}_{2}^{'})$ such that
$\bar{R}_{1}(D_{1},\mathcal{P}_{1}^{'},\mathcal{P}_{2}^{'})\geq
\bar{R}_{1}(D_{1},\mathcal{P}_{1}^*,\mathcal{P}_{2}^*),$
$\bar{R}_{2}(D_{1},\mathcal{P}_{1}^{'},\mathcal{P}_{2}^{'})\geq
\bar{R}_{2}(D_{1},\mathcal{P}_{1}^*,\mathcal{P}_{2}^*)$ and at
least one of these equalities is strict.
\end{defn}

Intuitively, this notion allows for eliminating Nash equilibria
which are dominated by other equilibrium points. One would expect
the rationality of the players to steer them away from such
dominated equilibria, and hence, they will ultimately settle in
one of the admissible points. This approach allows for modifying
the solution set for our low level game to only include admissible
Nash equilibria
$S^{*}(D_{1})=\{(\mathcal{P}_{1},\mathcal{P}_{2}):(\mathcal{P}_{1},\mathcal{P}_{2})\;
\text{is an admissible Nash equilibrium of the low level game}\}$.
The following result establishes the existence of a single
admissible Nash equilibrium in this set (for any choice of $D_1$)
\begin{thm}\label{thm:unique}
For any strategy $D_{1}$ of the base-station, and any channel
distribution function, there exists a \textbf{single} admissible
Nash equilibrium for the low level power/rate allocation game
(i.e., for any $D_{1}$, $S^{*}(D_{1})$ is a singleton).
\end{thm}
\begin{proof}
If $D_1$ is the same as the region given by the
Section~\ref{sec:basenp}, then the optimal solution is
time-sharing, and the Nash equilibrium is unique (as established
earlier). For other $D_1$, we establish uniqueness of the
admissible Nash equilibrium by contradiction.

We let $(\lambda_{1}^{*},\lambda_{2}^*)$ and
$(\lambda_{1}^{'},\lambda_{2}^{'})$ be the two pairs of
water-levels corresponding to equilibria. Then, by definition, the
two average power constraints are satisfied with equality with
these two pairs of water-levels, that is
$\bar{P}_{1}(\lambda_{1}^{*},\lambda_{2}^{*})=\bar{P}_{1},\bar{P}_{2}(\lambda_{1}^{*},\lambda_{2}^{*})=\bar{P}_{2}
$,
$\bar{P}_{1}(\lambda_{1}^{'},\lambda_{2}^{'})=\bar{P}_{1},\bar{P}_{2}(\lambda_{1}^{'},\lambda_{2}^{'})=\bar{P}_{2}
$. Noting that we {\bf are not} at a time sharing point, we claim:
\begin{enumerate}

\item If $\lambda_{1}^{*}=\lambda_{1}^{'}$, we have
$\lambda_{2}^{*}=\lambda_{2}^{'}$. If not, we will have
$\bar{P}_{1}(\lambda_{1}^{'},\lambda_{2}^{'})>\bar{P}_{1},\bar{P}_{2}(\lambda_{1}^{'},\lambda_{2}^{'})<\bar{P}_{2}
$ when $\lambda_{2}^{*}>\lambda_{2}^{'}$ and
$\bar{P}_{1}(\lambda_{1}^{'},\lambda_{2}^{'})<\bar{P}_{1},\bar{P}_{2}(\lambda_{1}^{'},\lambda_{2}^{'})>\bar{P}_{2}
$ when $\lambda_{2}^{*}<\lambda_{2}^{'}$. Thus we come to a
contradiction.

\item If $\lambda_{1}^{*}<\lambda_{1}^{'}$, we have
$\lambda_{2}^{*}<\lambda_{2}^{'}$. If not, we will have
$\bar{P}_{1}(\lambda_{1}^{'},\lambda_{2}^{'})>\bar{P}_{1},\bar{P}_{2}(\lambda_{1}^{'},\lambda_{2}^{'})<\bar{P}_{2}
$ when $\lambda_{2}^{*}\geq\lambda_{2}^{'}$. Thus we come to a
contradiction.

\item If $\lambda_{1}^{*}>\lambda_{1}^{'}$, we have
$\lambda_{2}^{*}>\lambda_{2}^{'}$. If not, we will have
$\bar{P}_{1}(\lambda_{1}^{'},\lambda_{2}^{'})<\bar{P}_{1},\bar{P}_{2}(\lambda_{1}^{'},\lambda_{2}^{'})>\bar{P}_{2}
$ when $\lambda_{2}^{*}\leq\lambda_{2}^{'}$. Thus we come to a
contradiction. \end{enumerate}

The two water-level pairs, therefore, have a strict order. We can
define the relationship $<$ for the water-level pairs and say
$(\lambda_{1}^{*},\lambda_{2}^{*})<(\lambda_{1}^{'},\lambda_{2}^{'})$,
if $\lambda_{1}^{*}<\lambda_{1}^{'}$ and
$\lambda_{2}^{*}<\lambda_{2}^{'}$. Suppose
$(\lambda_{1}^{*},\lambda_{2}^{*})<(\lambda_{1}^{'},\lambda_{2}^{'})$,
we claim that
$\bar{R}_{1}(D_{1},\mathcal{P}_{1}^{*},\mathcal{P}_{2}^{*})>\bar{R}_{1}(D_{1},\mathcal{P}_{1}^{'},\mathcal{P}_{2}^{'})$
and
$\bar{R}_{2}(D_{1},\mathcal{P}_{1}^{*},\mathcal{P}_{2}^{*})>\bar{R}_{2}(D_{1},\mathcal{P}_{1}^{'},\mathcal{P}_{2}^{'})$.
Without loss of generality, we only need to prove the first part.
To show this, we can see that the sum of the interference from
user $2$ and the background noise is
\begin{eqnarray}
N_{1}^{'}(\lambda_{2})=\sigma^2,\nonumber
\end{eqnarray}
if $(h_{1},h_{2})\in D_{1}^c$, and
\begin{eqnarray}
N_{1}^{'}(\lambda_{2})=\sigma^2+(\lambda_{2}h_{2}-\sigma^2)^+.
\end{eqnarray}
if $(h_{1},h_{2})\in D_{1}$.

Since our solution is not time sharing, we can see that
$N_{1}^{'}(\lambda_{2})$ is a decreasing function of
$\lambda_{2}$. Thus $\lambda_{2}^*<\lambda_{2}'$ implies that
$\bar{R}_{1}(D_{1},\mathcal{P}_{1}^{*},\mathcal{P}_{2}^{*})>\bar{R}_{1}(D_{1},\mathcal{P}_{1}^{'},\mathcal{P}_{2}^{'})$
and our claim is true.

This claim means that the achievable utility pairs also have
strict order, i.e., the smaller the water-filling pair, the larger
the utility pair. With this strict order relationship among the
achievable utilities at the Nash equilibria, the unique admissible
Nash equilibrium is achieved with the minimum water-level pair.
This completes the proof.
\end{proof}

An {\em explicit} approach for achieving the unique admissible
equilibrium in our game is for all the users to follow the
iterative algorithm used in the proof of Theorem~\ref{thm:nash}
and agree off-line on the {\bf convention} of starting the
iteration with $\lambda_1(1)=\lambda_2(1)=0$. This agreement is
clearly in the best interest of the two users, and hence, is
consistent with the selfish behavior assumption.

Now, we turn our attention to characterizing efficient
base-station strategies. In the following we use
$\mathcal{P}_{iD_{1}}$ to refer to the unique power control policy
of each user, under strategy $D_{1}$, at the admissible Nash
equilibrium. Here, we borrow the following definition
from~\cite{Basar}.

\begin{defn}
A strategy $D_{1}^{*}$ is called a Stackelberg equilibrium
strategy for a given $(\mu_{1},\mu_{2})$, if
\begin{eqnarray}
R^{*}&=&\mu_{1}\bar{R}_{1}(D_{1}^{*},\mathcal{P}_{1D_{1}^{*}},\mathcal{P}_{2D_{1}^{*}})+\mu_{2}\bar{R}_{2}(D_{1}^{*},\mathcal{P}_{1D_{1}^{*}},\mathcal{P}_{2D_{1}^{*}})\nonumber\\
&\geq&
\mu_{1}\bar{R}_{1}(D_{1},\mathcal{P}_{1D_{1}},\mathcal{P}_{2D_{1}})+\mu_{2}\bar{R}_{2}(D_{1},\mathcal{P}_{1D_{1}},\mathcal{P}_{2D_{1}}),
\end{eqnarray}
for all $D_{1}$. Moreover, for any $\epsilon>0$, a strategy
$D_{1,\epsilon}^{*}$ is called an $\epsilon$-Stackelberg strategy
if
\begin{equation}
\mu_{1}\bar{R}_{1}(D_{1,\epsilon}^{*},\mathcal{P}_{1D_{1,\epsilon}^{*}},\mathcal{P}_{2D_{1,\epsilon}^{*}})+\mu_{2}\bar{R}_{2}(D_{1,\epsilon}^{*},\mathcal{P}_{1D_{1,\epsilon}^{*}},\mathcal{P}_{2D_{1,\epsilon}^{*}})\geq
R^{*}-\epsilon.
\end{equation}

\end{defn}

\begin{cor}\label{cor:equili}
For every pair $(\mu_{1},\mu_{2}), 0\leq \mu_{1}<\infty, 0\leq
\mu_{2}<\infty$, an $\epsilon$-Stackelberg strategy exists.
\end{cor}
\begin{proof}
Based on Property 4.2 of~\cite{Basar}, the only thing we need to
prove is that $R^{*}$ is bounded. Define $R_{i}^{o}$ as the
average rate the $i$th user can get when the other user is absent,
then
\begin{equation}
R^{*}=\mu_{1}\bar{R}_{1}(D_{1}^{*},\mathcal{P}_{1D_{1}^{*}},\mathcal{P}_{2D_{1}^{*}})+\mu_{2}\bar{R}_{2}(D_{1}^{*},\mathcal{P}_{1D_{1}^{*}},\mathcal{P}_{2D_{1}^{*}})\leq
\mu_{1}R_{1}^{o}+\mu_{2}R_{2}^{o}.
\end{equation}
This completes the proof.
\end{proof}

Combining Theorem~\ref{thm:unique} and Corollary~\ref{cor:equili},
we see that the proposed Stackelberg game setup has a very
desirable structure. For any given vector $\muv$, the existence of
equilibrium is guaranteed and the optimal policy for every
rational multiple access user in the low level game is unique.
Therefore, the users will have no difficulty in deciding the power
and rate levels in a distributed way. The following result
characterizes the achievable performance of the proposed
Stackelberg game.

\begin{thm}\label{lem:three}
Let
$\mathcal{G}_s=\bigcup\limits_{D_{1}}\{(\bar{R}_{1}(D_{1},\mathcal{P}_{1D_{1}},\mathcal{P}_{2D_{1}}),\bar{R}_{2}(D_{1},\mathcal{P}_{1D_{1}},\mathcal{P}_{2D_{1}}))\}$.
Then, $\mathcal{G}_s$ includes the three boundary points
$CR_{1},CR_{2},SP$ of the capacity region $\mathcal{G}_{c}$.
However, $\mathcal{G}_s$ does not include any other boundary
points of $\mathcal{G}_{c}$.
\end{thm}
\begin{proof}
It is easy to verify that $CR_{1}$ can be achieved by setting
$D_{1}=\phi$, which means the base-station will always decode user
2's signal first. The corresponding policy for user $1$ is to
water-fill over the background noise, while the optimal policy for
user $2$ is also water-filling but over the sum of the
interference from user 1 and the background noise. This is exactly
the same as the centralized policy that achieves the boundary
point $CR_{1}$. Similarly $CR_{2}$ can be achieved by setting
$D_{1}^c=\phi$, and $SP$ can be achieved by setting $D_{1}$ as the
same region given in the Section~\ref{sec:basenp}.

Now suppose that $\mathcal{G}_s$ includes another boundary point
$(\bar{R}_{1b},\bar{R}_{2b})$. Without loss of generality, suppose
that at this point $\mu_{1}>\mu_{2}$ and the corresponding optimal
central policy is $\mathcal{P}_{b},\mathcal{R}_{b}$. The partition
region that achieves this point is given by $D_b$. The
corresponding admissible power control pair is
$\mathcal{P}_{1D_b},\mathcal{P}_{2D_b}$. It was shown
in~\cite{Tse} that the power control policy that achieves any
boundary point is unique. Thus if the partition $D_b$ achieves
this point, at any fading state $(h_{1},h_{2})$, we have
\begin{eqnarray}
\mathcal{P}_{1D_b}(h_{1},h_{2})&=&\mathcal{P}_{1,b}(h_{1},h_{2}),\nonumber\\
\mathcal{P}_{2D_b}(h_{1},h_{2})&=&\mathcal{P}_{2,b}(h_{1},h_{2}).
\end{eqnarray}

Then at any fading state, the capacity region pentagons formed by
these two policies are same, which is also shown on
figure~\ref{fig:pantegon}.
\begin{figure}[thb]
\centering
\includegraphics[width=0.5 \textwidth]{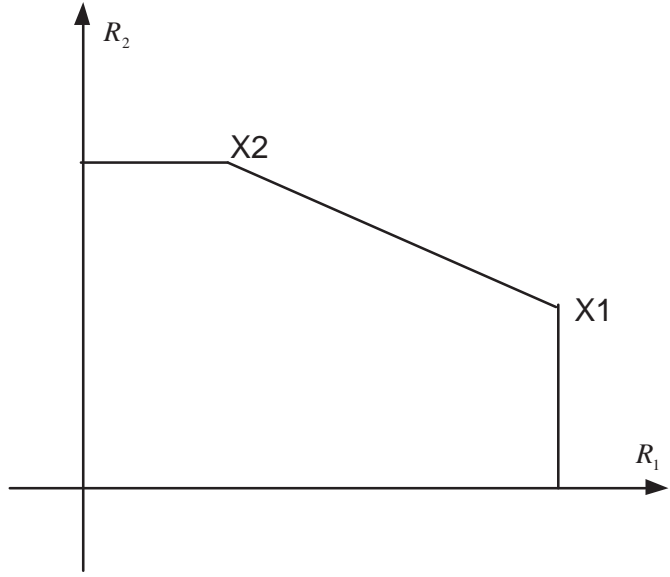}
\caption{The capacity region of the Gaussian multiple access
channel with fixed channel gains $(h_{1},h_{2})$.}
\label{fig:pantegon}
\end{figure}
For every fading state, the optimal rate control policy
$\mathcal{R}_{b}$ corresponds to the corner point $X{1}$. While
for the distributed power control, when $(h_{1},h_{2})\in D$, the
operating point is $X{2}$, and when $(h_{1},h_{2})\in D^c$, the
operating point is $X{1}$. Thus
\begin{eqnarray}
\bar{R}_{1}(D,\mathcal{P}_{1D},\mathcal{P}_{2D})&=&E_{\{\mathbf{h}\in
D\}}[R_{1,X{1}}(\mathbf{h})]+E_{\{\mathbf{h}\in
D^c\}}[R_{1,X{2}}(\mathbf{h})] \nonumber\\&<&E_{\{\mathbf{h}\in
D\}}[R_{1,X{1}}(\mathbf{h})]+E_{\{\mathbf{h}\in
D^c\}}[R_{1,X{1}}(\mathbf{h})]=\bar{R}_{1b},
\end{eqnarray}
which is a contradiction. This show the non-existence of $D$ that
achieves any other boundary point of the capacity region
$\mathcal{G}_{c}$.
\end{proof}

\begin{figure}[thb]
\centering
\includegraphics[width=0.5 \textwidth]{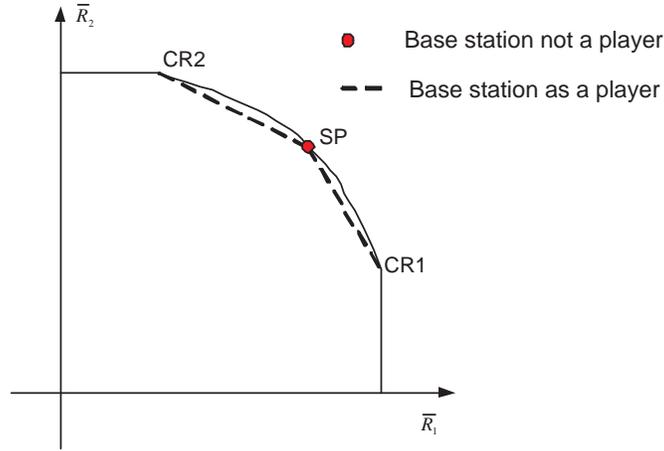}
\caption{The equilibria points of the Stackelberg power/rate
allocation game.} \label{fig:baseplayer}
\end{figure}

Theorem~\ref{lem:three} shows that the introduction of the
base-station as a leader of the game enlarged the achievable rate
region (as compared to the Nash game discussed earlier) but this
approach fails short of achieving the whole capacity region.
Figure~\ref{fig:baseplayer} compares the capacity region with the
Stackelberg achievable rate region assuming the following simple
base-station strategy: when $h_{1}\leq \alpha h_{2}$ the
base-station decodes user $1$ first and when $h_{1}\geq \alpha
h_{2}$ the base-station decodes user $2$ first. Under this
strategy, the rates at the Nash-equilibrium are:
\begin{eqnarray}
\bar{R}_{1}(\alpha)&=&
\int_{0}^{\infty}\int_{\frac{\sigma^2}{\lambda_{1}}+\frac{(\lambda_{2}h_{2}-\sigma^2)^{+}}{\lambda_{1}}}^{\alpha
h_{2}}
\frac{1}{2}\log_{2}\bigg(1+\frac{\lambda_{1}h_{1}-\sigma^2-(\lambda_{2}h_{2}-\sigma^2)^{+}}
{\sigma^2+(\lambda_{2}h_{2}-\sigma^2)^{+}}\bigg)f(h_{1},h_{2})dh_{1}dh_{2}\nonumber\\
&+&\int_{0}^{\infty}\int_{\max\{\alpha
h_{2},\frac{\sigma^2}{\lambda_{1}}\}}^{\infty}
\frac{1}{2}\log_{2}\bigg(1+\frac{\lambda_{1}h_{1}-\sigma^{2}}{\sigma^{2}}\bigg)f(h_{1},h_{2})dh_{1}dh_{2},
\end{eqnarray}

\begin{eqnarray}
\bar{R}_{2}(\alpha)&=&
\int_{0}^{\infty}\int_{\frac{\sigma^2}{\lambda_{2}}+\frac{(\lambda_{1}h_{1}-\sigma^2)^{+}}{\lambda_{2}}}^{\frac{h_{1}}{\alpha}
}
\frac{1}{2}\log_{2}\bigg(1+\frac{\lambda_{2}h_{2}-\sigma^2-(\lambda_{1}h_{1}-\sigma^2)^{+}}
{\sigma^2+(\lambda_{1}h_{1}-\sigma^2)^{+}}\bigg)f(h_{1},h_{2})dh_{1}dh_{2}\nonumber\\
&+&\int_{0}^{\infty}\int_{\max\{\frac{h_{1}}{\alpha}
,\frac{\sigma^2}{\lambda_{2}}\}}^{\infty}
\frac{1}{2}\log_{2}\bigg(1+\frac{\lambda_{2}h_{2}-\sigma^{2}}{\sigma^{2}}\bigg)f(h_{1},h_{2})dh_{1}dh_{2},
\end{eqnarray}
where $\lambda_{1},\lambda_{2}$ are the solutions to the following
equations:
\begin{eqnarray}
&&\int_{0}^{\infty}\int_{\frac{\sigma^2}{\lambda_{1}}+\frac{(\lambda_{2}h_{2}-\sigma^2)^{+}}{\lambda_{1}}}^{\alpha
h_{2}}
\bigg(\lambda_{1}-\frac{\sigma^2+(\lambda_{2}h_{2}-\sigma^2)^{+}}{h_{1}}\bigg)f(h_{1},h_{2})dh_{1}dh_{2}\nonumber\\
&&+\int_{0}^{\infty}\int_{\max\{\alpha h_{2}
,\frac{\sigma^2}{\lambda_{1}}\}}^{\infty}
\bigg(\lambda_{1}-\frac{\sigma^2}{h_{1}}\bigg)f(h_{1},h_{2})dh_{1}dh_{2}=\bar{P}_{1},\nonumber\\
&&\int_{0}^{\infty}\int_{\frac{\sigma^2}{\lambda_{2}}+\frac{(\lambda_{1}h_{1}-\sigma^2)^{+}}{\lambda_{2}}}^{\frac{h_{1}}{\alpha}
}
\bigg(\lambda_{2}-\frac{\sigma^2+(\lambda_{1}h_{1}-\sigma^2)^{+}}{h_{2}}\bigg)f(h_{1},h_{2})dh_{1}dh_{2}\nonumber\\
&&+\int_{0}^{\infty}\int_{\max\{\frac{h_{1}}{\alpha}
,\frac{\sigma^2}{\lambda_{2}}\}}^{\infty}
\bigg(\lambda_{2}-\frac{\sigma^2}{h_{2}}\bigg)f(h_{1},h_{2})dh_{1}dh_{2}=\bar{P}_{2}.
\end{eqnarray}

It is easy to verify that $CR_1$ is achieved by setting
$\alpha=0$, $CR_2$ is achieved by setting $\alpha=\infty$, and
$SP$ is achieved by setting
$\alpha=\frac{\lambda_{2}^o}{\lambda_{1}^o}$, where
$\lambda_{1}^o,\lambda_{2}^o$ are the water-filling levels given
in the Section~\ref{sec:basenp}. One can also prove the following
statement.
\begin{cor}
For the base-station that adopts the simple region partition
strategy, there always exists a Stackelberg equilibrium solution
for any pair $(\mu_{1},\mu_{2})$, if the channel gains are bounded
and satisfy $\min(h_{1})>0,\min(h_{2})>0$.
\end{cor}
\begin{proof}
Since $(h_{1},h_{2})$ are bounded, and
$\min(h_{1})>0,\min(h_{2})>0$, then
$\alpha\in[\min(h_{1})/\max(h_{2}),$ $\max(h_{1})/\min(h_{2})]$ is
a compact set. And for every $\alpha$, we have proved in
Theorem~\ref{thm:unique}, $S^{*}(\alpha)$ is a singleton, thus
based on~\cite{Basar}, for any pair $(\mu_{1},\mu_{2})$, there
exists a Stackelberg equilibrium solution.
\end{proof}

\subsection{Repeated Game Formulation}\label{repeated}
The inability of our Stackelberg game to achieve all the boundary
points of the capacity region can be attributed to the structural
difference between our successive decoding strategy and the
optimal decoding strategy characterized in \cite{Tse}. In
particular, the optimal decoding strategy will always decode user
$1$ first (i.e., for all channel states) if $\mu_1<\mu_2$, whereas
in our formulation the decoding order is a function of the channel
state. Unfortunately, if we adopt any {\em static} decoding order,
the game will always settle at one of the corner points of the
capacity region as argued in the previous section. To solve this
problem, we pursue our last resort of replacing the static game
formulation with a dynamic one.

The static formulation assumes that the players interact with each
other only once. This assumption models the case where the
topology of the network changes quickly. In a more slowly varying
environments, a dynamic game formulation seems more appropriate.
Specifically, we call a game where the players interact for $T>1$
instances a dynamic game\footnote{We note that every game stage is
assumed long enough to justify invoking the ergodic assumption
within every stage.}. An example of a dynamic game is the repeated
game where the same static game is played many times. Obviously,
the users can play this game by repeating the same static
strategy~\cite{mitgame}. But, the advantage of the repeated game
framework is that the players can do better than just repeating
the same static strategy. The idea is that, since the players will
interact with each other many times, they can learn each other's
strategies, which may allow them to cooperate to obtain higher
payoffs. In this case, the players can start cooperating and if
one player deviates from the cooperation phase, the other players
will adjust their strategies to punish the deviating player. The
punishment {\em threat} is credible only if the deviating player
achieves a lower payoff under punishment as compared with the
cooperating phase. Under these circumstances, the users will have
no desire to deviate from the cooperation phase, thus all the
users can achieve higher utilities as compared to the static
scenario.

In the repeated game, the utility of each player can be defined as
as a discounted sum of the payoff achieved in each stage. We refer
to the discount factor by $\delta$, where $0<\delta<1$. The larger
$\delta$ is the more patient the player is. In the proof of the
following theorem, we use a generalized version of a result due to
Aumann and Shapley~\cite{mitgame}~\cite{Aumann} and define the
payoff of the repeated game as the time-average of payoff at each
stage.
\begin{thm}\label{thm:reb}
As $T\rightarrow \infty$, all the boundary points of the capacity
region are achievable under the repeated game setup with the
base-station as the game leader. Moreover, the corresponding
equilibria are subgame perfect.
\end{thm}
\begin{proof}
In order to prove our claims, we need to construct a subgame
perfect strategy that achieves every boundary point. Consider the
following strategy: The base-station announces its rate award
vector $\muv$, then the game proceeds in the following way:
\begin{enumerate}
\item $t=1$, each user uses the optimal centralized control policy
$\mathcal{P}_{c}$ and rate control policy $\mathcal{R}_{c}$ that
maximize $\sum\mu_{i}\bar{R}_{i}$. Under this point, each user
gets a rate $\bar{R}_{i}$.

\item if user $1$ deviates from the centralized control policy at
stage $t=t_{d}$, then the base-station will punish user $1$ by
moving to the corner point $CR_{2}$ for $T_{1}$ periods (i.e.,
decoding user $1$ first for $T_1$ stages). The parameter $T_{1}$
is chosen such that
\begin{equation}
\bar{R}_{1,CR_1}+\sum_{i=1}^{T_{1}}\bar{R}_{1,CR_2}<\sum_{i=1}^{T_{1}}\bar{R}_{1}.
\end{equation}
After $T_{1}$ periods, the players return to the cooperative
phase. If user $2$ deviates, the base-station can also punish it
for $T_{2}$ phases, which can be chosen in a similar way, by
moving to the corner point $CR_{1}$.
\end{enumerate}
The conditions on $T_{i}$ ensures that any gain obtained from
deviating is removed at the punishment phase, so no sequence of a
finite or infinite number of deviations can increase user $i$'s
payoff. Moreover, although it is costly for the base-station to
carry out the punishment, any finite number of such losses are
costless in the long run. This proves the subgame perfection of
the strategy.
\end{proof}

\subsection{Arbitrary Number of Users}\label{arbitrary}
All our results generalize naturally to the $N$ user channel
except for Theorem~\ref{thm:unique}. The arguments used in the
proof do not carry over for $N>3$, and hence, we can not guarantee
the uniqueness of the admissible Nash equilibrium. However, if the
multiple-access users choose the Nash equilibrium corresponding to
the iterative algorithm used in the proof with
${\lambdav=\mathbf{0}}$, then the rest of our results in
Section~\ref{stack} hold.  The base-station can announce this
initial condition in the first stage of the Stackelberg game. All
users will be forced to follow this strategy since any deviation
can result in the catastrophic event of unsuccessful decoding.

For the sake of completeness, we detail in this section the
generalization of our Nash game. The other scenarios follow
virtually the same lines, and hence, are omitted for brevity. We
first restate our assumption that all the users are informed {\em
a-priori} of all the CSI. This is exactly the same assumption used
in \cite{CemNa,Farhad,Xiao,Zhou,Alpcan,Sung}, and corresponds to a
game with complete information. In the Nash formulation, every
user treats the signals from other users as noise. The optimal
power control policy of each user is to water-fill over the sum of
the interference and the background noise, i.e.,
 \begin{eqnarray}
 \mathcal{P}_{i}(\mathbf{h})=\Bigg(\lambda_{i}-\frac{\sigma^2+\sum\limits_{j=1,j\neq
 i}^{N}h_{j}\mathcal{P}_{j}(\mathbf{h})}{h_{i}}\Bigg)^+.
 \end{eqnarray}
Each user will adjust its water level depending on the levels of
the other users. At the Nash equilibrium points the water levels
$\lambda_{i},\;i=1,\cdots,N$ satisfy all power constraints with
equality. In order to show that the only Nash equilibrium of this
game is the maximum sum-rate point, we generalize the proof of
Theorem~\ref{thm:nashsum}. In particular, we show that at the
equilibrium only one user will transmit at any fading state then
it is easy to verify that the power control policy of each user at
the equilibrium is exactly the same as the corresponding central
policy for the point $SP$. Without loss of generality, suppose
that users $1$ to $M$ are transmitting simultaneously at certain
fading states, then for each transmitting user, we have
\begin{eqnarray}
\mathcal{P}_{1}+\frac{\sigma^2+\sum\limits_{j=2}^{M}h_{j}\mathcal{P}_{j}}{h_{1}}&=&\lambda_{1},\nonumber\\
&\cdots&\nonumber\\
\mathcal{P}_{i}+\frac{\sigma^2+\sum\limits_{j=1,j\neq i}^{M}h_{j}\mathcal{P}_{j}}{h_{i}}&=&\lambda_{i},\\
&\cdots&\nonumber\\
\mathcal{P}_{M}+\frac{\sigma^2+\sum\limits_{j=1}^{M-1}h_{j}\mathcal{P}_{j}}{h_{M}}&=&\lambda_{M}.\nonumber
\end{eqnarray}
These conditions imply that
$\lambda_{i}h_{i}=\lambda_{j}h_{j},\forall i,j=1,\cdots,M$. With
continuous probability density functions, this happens with
probability zero. Then with probability one, at any fading state
only one user will transmit. If user $i$ transmits, the sum of
background noise and the signal of user $i$ should be larger than
the water level of user $j$, and hence, $h_{i}$ should satisfy
\begin{eqnarray}
\bigg(\lambda_{i}-\frac{\sigma^2}{h_{i}}\bigg)\frac{h_{i}}{h_{j}}+\frac{\sigma^2}{h_{j}}=\lambda_{i}\frac{h_{i}}{h_{j}}\geq
\lambda_{j},\forall j\neq i.
\end{eqnarray}

\section{Vector Channels}\label{vector}

Thus far, we have presented our results for the scalar channel
where the base-station is only equipped with one receive antenna.
In this section, we extend our study to the vector multiple access
channel where the base-station is equipped with $N_{r}$ receive
antennas. Our goal is to see if our previous conclusions carry
through or not. Again to simplify the presentation, we focus on
the two user scenario. The signal received at any time $n$ is
given by
\begin{equation}
\mathbf{y}(n)=\sum\limits_{i=1}^2
\mathbf{h}_{i}(n)x_{i}(n)+\mathbf{z}(n),
\end{equation}
where
$\mathbf{h}_{i}(n)=[\sqrt{h_{1i}},\sqrt{h_{2i}},\cdots,\sqrt{h_{N_{r}i}}]^T$
is the $N_{r}\times 1$ fading vector from user $i$ to the $N_{r}$
receive antennas. As before, we assume that the fading processes
have a joint continuous distribution with a bounded density.
$\mathbf{z}(n)$ is the gaussian noise vector at the $N_{r}$
receive antenna with correlation matrix
$E[\mathbf{z}\mathbf{z}^T]=\sigma^2\mathbf{I}_{Nr}$.

Similar to the scalar channel case, we first consider the static
Nash formulation where the only players of the game are the
multiple access users. The strategy space of user $i$ is still
$\mathcal{F}_{i}=\{\mathcal{P}_{i}:E_{\mathbf{H}}[\mathcal{P}_{i}]\leq\bar{P}_{i},
\mathcal{P}_{i}(\mathbf{H})\geq 0\}$ with
$\mathbf{H}=[\mathbf{h}_{1},\mathbf{h}_{2}]$. The payoff function
of user $i$ is still the average achievable rate
$\bar{R}_{i}=E_{\mathbf{H}}[\mathcal{R}_{i}]$. It is easy to see
that for any power control strategy
$\mathcal{P}_{2}(\mathbf{h}_{1},\mathbf{h}_{2})$ of user $2$, the
optimal power control policy of user $1$ is the solution to the
following optimization problem
\begin{eqnarray}\label{equ:mimoop}
\max \limits_{\mathcal{P}_{1}}\quad
\bar{R}_{1}&=&E_{\mathbf{H}}\bigg[\frac{1}{2}\log_{2}\bigg(
\det\Big(\sigma^2\mathbf{I}_{N_{r}}+\mathcal{P}_{1}(\mathbf{h}_{1},\mathbf{h}_{2})\mathbf{h}_{1}\mathbf{h}_{1}^T
+\mathcal{P}_{2}(\mathbf{h}_{1},\mathbf{h}_{2})\mathbf{h}_{2}\mathbf{h}_{2}^T\Big)\bigg)\nonumber\\
&-&\frac{1}{2}\log_{2}\bigg( \det\Big(\sigma^2\mathbf{I}_{N_{r}}
+\mathcal{P}_{2}(\mathbf{h}_{1},\mathbf{h}_{2})\mathbf{h}_{2}\mathbf{h}_{2}^T\Big)\bigg)\bigg],\nonumber\\
&\text{s.t.}&\quad
\mathcal{P}_{1}(\mathbf{h}_{1},\mathbf{h}_{2})\in \mathcal{F}_{1}.
\end{eqnarray}
Given any power control strategy
$\mathcal{P}_{1}(\mathbf{h}_{1},\mathbf{h}_{2})$ of user $1$, the
optimal power control strategy of user $2$ is a solution to a
similar problem. The difference between the vector and scalar
channels is highlighted in the following result.

\begin{thm}\label{thm:mimonash}
There exists a unique Nash equilibrium for the distributed
power/rate allocation game in the vector multiple access channel.
At this equilibrium, the power control policy of each user is the
same as the central policy that achieves the maximum sum-rate
point $SP$. The achieved rates, however, are strictly smaller than
the rates corresponding to $SP$.
\end{thm}
\begin{proof}
Given the power control policy
$\mathcal{P}_{2}(\mathbf{h}_{1},\mathbf{h}_{2})$, it is easy to
see that $E_{\mathbf{H}}\bigg[\frac{1}{2}\log_{2}\bigg(
\det\Big(\sigma^2\mathbf{I}_{N_{r}}
+\mathcal{P}_{2}(\mathbf{h}_{1},\mathbf{h}_{2})\mathbf{h}_{2}\mathbf{h}_{2}^T\Big)\bigg)\bigg]$
is a constant, thus the solution to the optimization
problem~\eqref{equ:mimoop} is the same as the solution to the
following optimization problem
\begin{eqnarray}
\max\limits_{\mathcal{P}_{1}}\quad f(\mathcal{P}_{1})
&=&E_{\mathbf{H}}\bigg[\frac{1}{2}\log_{2}\bigg(
\det\Big(\sigma^2\mathbf{I}_{N_{r}}+\mathcal{P}_{1}(\mathbf{h}_{1},\mathbf{h}_{2})\mathbf{h}_{1}\mathbf{h}_{1}^T
+\mathcal{P}_{2}(\mathbf{h}_{1},\mathbf{h}_{2})\mathbf{h}_{2}\mathbf{h}_{2}^T\Big)\bigg)\bigg],\nonumber\\
&&\text{s.t.}\quad
\mathcal{P}_{1}(\mathbf{h}_{1},\mathbf{h}_{2})\in \mathcal{F}_{1}.
\end{eqnarray}

Since
 $\sigma^2\mathbf{I}_{N_{r}}+\mathcal{P}_{1}(\mathbf{h}_{1},\mathbf{h}_{2})\mathbf{h}_{1}\mathbf{h}_{1}^T
+\mathcal{P}_{2}(\mathbf{h}_{1},\mathbf{h}_{2})\mathbf{h}_{2}\mathbf{h}_{2}^T$
is positive definite, and the $\log_{2}(\det(.))$ function is
concave in the set of positive definite matrices, then the
objective function is concave in the set of power allocation
policies. The constraint set is convex and it is easy to verify
that the Slater's condition is satisfied. Hence, there exists a
constant $\gamma_{1}$, such that the solution
to~\eqref{equ:mimoop} is the same as the solution to the following
optimization problem:
\begin{eqnarray}
\max \limits_{\mathcal{P}_{1}}\quad
L_{1}(\mathcal{P}_{1}(\mathbf{h}_{1},\mathbf{h}_{2}),\gamma_{1})&=&E_{\mathbf{H}}\bigg[\frac{1}{2}\log_{2}\bigg(
\det\Big(\sigma^2\mathbf{I}_{N_{r}}+\mathcal{P}_{1}(\mathbf{h}_{1},\mathbf{h}_{2})\mathbf{h}_{1}\mathbf{h}_{1}^T
+\mathcal{P}_{2}(\mathbf{h}_{1},\mathbf{h}_{2})\mathbf{h}_{2}\mathbf{h}_{2}^T\Big)\bigg)\bigg]\nonumber\\&-&\gamma_{1}E_{\mathbf{H}}[\mathcal{P}_{1}(\mathbf{h}_{1},\mathbf{h}_{2})]
\end{eqnarray}

The KKT necessary and sufficient conditions of this optimization
problem is
\begin{eqnarray}\label{equ:mimop1}
\frac{\partial L_{1}}{\partial \mathcal{P}_{1}}=\mathbf{h}_{1}^T
\bigg(\sigma^2\mathbf{I}_{N_{r}}+\mathcal{P}_{1}(\mathbf{h}_{1},\mathbf{h}_{2})\mathbf{h}_{1}\mathbf{h}_{1}^T
+\mathcal{P}_{2}(\mathbf{h}_{1},\mathbf{h}_{2})\mathbf{h}_{2}\mathbf{h}_{2}^T\bigg)^{-1}\mathbf{h}_{1}-\gamma_{1}&=&0. \nonumber\\
\gamma_{1}&\geq& 0.
\end{eqnarray}

Using the matrix inversion lemma~\cite{pramod}
\begin{eqnarray}\label{equ:matrixinv}
(\mathbf{A}+\mathbf{x}\mathbf{x}^t)^{-1}=\mathbf{A}^{-1}-\frac{\mathbf{A}^{-1}\mathbf{x}\mathbf{x}^t\mathbf{A}^{-1}}{1+\mathbf{x}^t\mathbf{A}^{-1}\mathbf{x}},
\end{eqnarray}
we come to
\begin{eqnarray}
\frac{\partial L_{1}}{\partial
\mathcal{P}_{1}}=\frac{\mathbf{h}_{1}^{T}\bigg(\sigma^2\mathbf{I}_{N_{r}}+\mathcal{P}_{2}(\mathbf{h}_{1},\mathbf{h}_{2})\mathbf{h}_{2}\mathbf{h}_{2}^T\bigg)^{-1}\mathbf{h}_{1}}
{1+\mathbf{h}_{1}^T\bigg(\sigma^2\mathbf{I}_{N_{r}}+\mathcal{P}_{2}(\mathbf{h}_{1},\mathbf{h}_{2})\mathbf{h}_{2}\mathbf{h}_{2}^T\bigg)^{-1}\mathbf{h}_{1}\mathcal{P}_{1}(\mathbf{h}_{1},\mathbf{h}_{2})}-\gamma_{1}&=&0,\\\nonumber
\gamma_{1}&\geq&0.
\end{eqnarray}
Considering the condition
$\mathcal{P}_{1}(\mathbf{h}_{1},\mathbf{h}_{2})\geq0$, we get
\begin{eqnarray}
\mathcal{P}_{1}(\mathbf{h}_{1},\mathbf{h}_{2})=\Bigg(\lambda_{1}-\frac{1}{\mathbf{h}_{1}^{T}\bigg(\sigma^2\mathbf{I}_{N_{r}}+\mathcal{P}_{2}(\mathbf{h}_{1},\mathbf{h}_{2})\mathbf{h}_{2}\mathbf{h}_{2}^T\bigg)^{-1}\mathbf{h}_{1}}\Bigg)^{+},
\end{eqnarray}
where $\lambda_{1}=\frac{1}{\gamma_{1}}$ is a constant that
satisfies the average power constraint of user $1$ with equality.
Similarly, given $\mathcal{P}_{1}(\mathbf{h}_{1},\mathbf{h}_{2})$,
we get the following optimality condition
\begin{eqnarray}\label{equ:mimop2}
\mathbf{h}_{2}^T
\bigg(\sigma^2\mathbf{I}_{N_{r}}+\mathcal{P}_{1}(\mathbf{h}_{1},\mathbf{h}_{2})\mathbf{h}_{1}\mathbf{h}_{1}^T
+\mathcal{P}_{2}(\mathbf{h}_{1},\mathbf{h}_{2})\mathbf{h}_{2}\mathbf{h}_{2}^T\bigg)^{-1}\mathbf{h}_{2}-\gamma_{2}&=&0,\nonumber\\
\gamma_{2}&\geq&0.
\end{eqnarray}
The optimal policy of user $2$ is therefore
\begin{eqnarray}
\mathcal{P}_{2}(\mathbf{h}_{1},\mathbf{h}_{2})=\Bigg(\lambda_{2}-\frac{1}{\mathbf{h}_{2}^{T}\bigg(\sigma^2\mathbf{I}_{N_{r}}+\mathcal{P}_{1}(\mathbf{h}_{1},\mathbf{h}_{2})\mathbf{h}_{1}\mathbf{h}_{1}^T\bigg)^{-1}\mathbf{h}_{2}}\Bigg)^{+},
\end{eqnarray}
where $\lambda_{2}$ is the constant that satisfies the average
power constraint of user $2$ with equality. Applying the results
of~\cite{pramod} to the fading multiple access channel with
$N_{r}$ receive antennas, we know
that~\eqref{equ:mimop1}~\eqref{equ:mimop2} are exactly the
optimality conditions for the following optimization problem
\begin{eqnarray}\label{equ:mimosum}
\max\limits_{\mathcal{P}_{1},\mathcal{P}_{2}}\quad
\bar{R}_{sum}(\mathcal{P}_1,\mathcal{P}_{2})&=&E_{\mathbf{H}}[\mathcal{R}_{1}+\mathcal{R}_{2}]\nonumber\\&=&E_{\mathbf{H}}\bigg[\frac{1}{2}\log_{2}\bigg(
\det\Big(\mathbf{I}_{N_{r}}+\frac{\mathcal{P}_{1}(\mathbf{h}_{1},\mathbf{h}_{2})\mathbf{h}_{1}\mathbf{h}_{1}^T
+\mathcal{P}_{2}(\mathbf{h}_{1},\mathbf{h}_{2})\mathbf{h}_{2}\mathbf{h}_{2}^T}{\sigma^2}\Big)\bigg)\bigg],\nonumber\\
&&s.t. \quad \mathcal{P}_{1}(\mathbf{h}_{1},\mathbf{h}_{2})\in
\mathcal{F}_{1}, \mathcal{P}_{2}(\mathbf{h}_{1},\mathbf{h}_{2})\in
\mathcal{F}_{2}.
\end{eqnarray}

One can easily verify that the optimization
problem~\eqref{equ:mimosum} will maximize the sum-rate at the
base-station. This means the optimal policy of each user aiming to
maximize its \textbf{own} rate while treating the signal of the
other user as interference is exactly the same as the power
control policy that maximizes the sum-rate at the base-station. A
similar observation has been made in the Gaussian multiple access
channel in~\cite{weiyumimo}.

Therefore, we can apply the following iterative process to get the
power control policy at the Nash equilibrium point. Starting at
$\mathcal{P}_{1}=0,\mathcal{P}_{2}=0$, each user takes a turn to
water-fill over the combined interference and the background
noise. At each step, the objective function of~\eqref{equ:mimosum}
increases. But with limited average power at the users, the
objective function~\eqref{equ:mimosum} has an upper-bound. Thus,
this process will converge, which means the Nash equilibrium
exists. At the convergence point, the optimality
conditions~\eqref{equ:mimop1}~\eqref{equ:mimop2} hold, which means
the power control policy of each user at the Nash equilibrium is
the same as the optimal policy that maximizes the sum-rate at the
base-station. The uniqueness of the power control policy that
maximizes the sum-rate~\cite{pramod} implies the uniqueness of the
Nash equilibrium point. This proves our first two claims.

From~\cite{pramod}, we know the optimal central control policy is
not time-sharing. Hence, in some channel fading states, the
transmission power of both users will be larger than zero. In
these cases, the capacity region pentagon is shown in
Figure~\ref{fig:mimopolygon}.
\begin{figure}[thb]
\centering
\includegraphics[width=0.5 \textwidth]{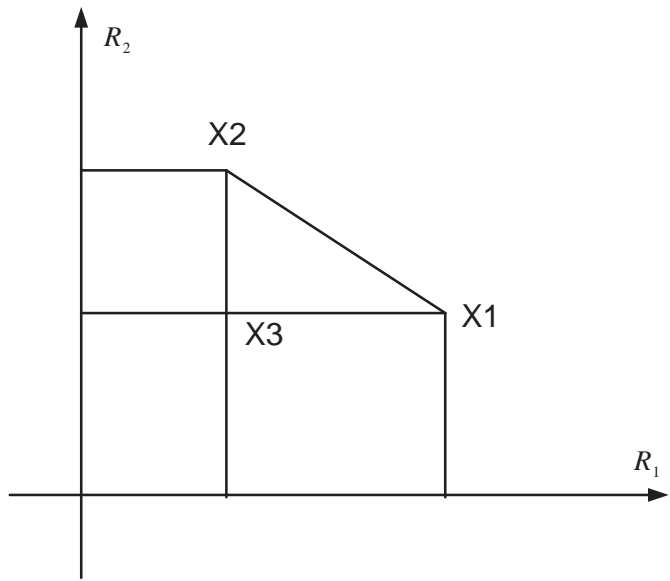}
\caption{The capacity region pentagon for fixed channel gains.}
\label{fig:mimopolygon}
\end{figure}
We can easily see that the central rate control policy will always
operate on one of the boundary points (the line between $X1$ and
$X2$), but the distributed scheme will always choose the point
$X3$. We have either
\begin{eqnarray}
E_{\mathbf{H}}[R_{1N}]<E_{\mathbf{H}}[R_{1,sum}]
\end{eqnarray}
or
\begin{eqnarray}
E_{\mathbf{H}}[R_{2N}]<E_{\mathbf{H}}[R_{2,sum}].
\end{eqnarray}
This completes the proof.
\end{proof}

Theorem~\ref{thm:mimonash} contrasts the scalar scenario, where
the Nash equilibrium rate is the same as the maximum sum-rate. The
reason is that in the scalar multiple-access channel, the strategy
that maximizes the sum-rate is time sharing. In the vector case,
on the other hand, we have $\min(N,N_r)$ degrees of freedom, and
hence, more than one user are allowed to transmit at any fading
state. The central control policy will choose to operate at one of
the boundary points, but because of the interference, the multiple
access users will distributively choose a point that is strictly
inside the capacity region at the Nash equilibrium point.

Our Stackelberg game can also be extended to the vector multiple
access channel. Similar to the scalar case, the base-station
partitions the space of $(\mathbf{h}_{1},\mathbf{h}_{2})$ into two
region $D_{1},D_{1}^c$, and decodes user $1$ first in $D_{1}$ and
decode user $2$ first in the region $D_{1}^c$. The following
results do not depend on the specific choice of $D_{1}$. The
strategy space of user $i$ is still $\mathcal{F}_{i}$, and the
payoff function of each user is still the supremum of achievable
average rate.
\begin{thm}
There exists a unique admissible Nash equilibrium for the low
level game. The Stackelberg game achieves the two corner points of
the capacity region but doesn't achieve the maximum sum-rate
point.
\end{thm}
\begin{proof}
The proof of the existence of a unique admissible Nash equilibrium
under any base-station strategy follows essentially the same lines
as the proofs of Theorems~\ref{thm:nash} and~\ref{thm:unique}. The
only additional requirement is to prove that
$\mathcal{P}_{1}(\mathbf{h}_{1},\mathbf{h}_{2})$ is a
non-decreasing function of $\lambda_{1}$ and a non-increasing
function of $\lambda_{2}$.

Based on the proof of Theorem~\ref{thm:mimonash}, we know that the
optimal power control policy of user $1$ is
\begin{eqnarray}
\mathcal{P}_{1}(\mathbf{h}_{1},\mathbf{h}_{2})&=&\Bigg(\lambda_{1}-\frac{\sigma^2}{\parallel\mathbf{h}_{1}\parallel^2}\Bigg)^{+},\quad\text{if}\quad
(\mathbf{h}_{1},\mathbf{h}_{2})\in D_{1}^c,\nonumber\\
\mathcal{P}_{1}(\mathbf{h}_{1},\mathbf{h}_{2})&=&\Bigg(\lambda_{1}-\frac{1}{\mathbf{h}_{1}^{T}\bigg(\sigma^2\mathbf{I}_{N_{r}}+\mathbf{h}_{2}\mathbf{h}_{2}^T\Big(\lambda_{2}-\frac{\sigma^2}{\parallel\mathbf{h}_{2}\parallel^{2}}\Big)^{+}\bigg)^{-1}\mathbf{h}_{1}}\Bigg)^{+},
\quad\text{if}\quad (\mathbf{h}_{1},\mathbf{h}_{2})\in D_{1}.
\end{eqnarray}

It is easy to verify that
$\mathcal{P}_{1}(\mathbf{h}_{1},\mathbf{h}_{2})$ is a
non-decreasing function of $\lambda_{1}$. To show that
$\mathcal{P}_{1}(\mathbf{h}_{1},\mathbf{h}_{2})$ is a
non-increasing function of $\lambda_{2}$, we only need to show
that
$\mathbf{h}_{1}^{T}\bigg(\sigma^2\mathbf{I}_{N_{r}}+\mathbf{h}_{2}\mathbf{h}_{2}^T\Big(\lambda_{2}-\frac{\sigma^2}{\parallel\mathbf{h}_{2}\parallel^{2}}\Big)^{+}\bigg)^{-1}\mathbf{h}_{1}$
is a non-increasing function of $\lambda_{2}$.

Using the matrix inversion lemma~\eqref{equ:matrixinv}, we have
\begin{eqnarray}
\mathbf{h}_{1}^{T}\bigg(\sigma^2\mathbf{I}_{N_{r}}+\mathbf{h}_{2}\mathbf{h}_{2}^T(\lambda_{2}-\frac{\sigma^2}
{\parallel\mathbf{h}_{2}\parallel^{2}})^{+}\bigg)^{-1}\mathbf{h}_{1}
&=&\mathbf{h}_{1}^{T}\bigg(\frac{\mathbf{I}_{N_{r}}}{\sigma^2}-\frac{\Big(\lambda_{2}-\frac{\sigma^2}{\parallel\mathbf{h}_{2}\parallel^{2}}\Big)^{+}\mathbf{h}_{2}\mathbf{h}_{2}^T}
{\sigma^4+\sigma^2\Big(\parallel\mathbf{h}_{2}\parallel^{2}\lambda_{2}-\sigma^2\Big)^{+}}\bigg)\mathbf{h}_{1}\nonumber\\
&=&\frac{\parallel\mathbf{h}_{1}\parallel^2}{\sigma^2}-\mid\mathbf{h}_{2}^T\mathbf{h}_{1}\mid^2g(\lambda_{2}),
\end{eqnarray}
in which
\begin{eqnarray}
g(\lambda_{2})=\frac{\Big(\lambda_{2}-\frac{\sigma^2}{\parallel\mathbf{h}_{2}\parallel^{2}}\Big)^{+}}
{\sigma^4+\sigma^2\Big(\parallel\mathbf{h}_{2}\parallel^2\lambda_{2}-\sigma^2\Big)^{+}}.
\end{eqnarray}

It is easy to verify that $g(\lambda_{2})$ is a non-decreasing
function of $\lambda_{2}$, thus we come to the conclusion that
$\mathcal{P}_{1}(\mathbf{h}_{1},\mathbf{h}_{2})$ is a
non-decreasing function of $\lambda_{1}$ and a non-increasing
function of $\lambda_{2}$. To achieve the corner points, the
base-station can just set $D_{1}$ to be the whole set, in one
case, and the empty set in the other case.

We prove the nonexistence of a base-station strategy that achieves
the sum-rate point by contradiction. Suppose that a partition
$D_{1}$ achieves the sum-rate point. Since the unique power
control policy that achieves the maximum sum-rate point is to
water-fill over the sum of the interference and the background
noise for {\bf both users}, then in the region $D_{1}$, user $1$
should stop sending. Because in this region, the optimal
distributed power control policy of user $2$ is to water-fill only
over the background noise. Similarly, in the region $D_{1}^c$,
user $2$ should also stop sending. Then we come to a time-sharing
solution, which cannot achieve the maximum sum-rate point and we
have our contradiction.
\end{proof}

Finally, if the users have the opportunity to interact many times
then any boundary point of the capacity region of the vector
multiple access channel can be achieved as a subgame perfect
equilibrium. Moreover, the users can use the same strategies
developed in Theorem~\ref{thm:reb} to achieve these boundary
points.

\section{Conclusions}\label{conclusion}
This paper has developed a game theoretic framework for
distributed resource allocation in fading multiple access
channels. In our first result, we showed that the opportunistic
communications principle can be obtained as the unique Nash
equilibrium of a water-filling game. By introducing the
base-station as a player, we were able to achieve all the corner
points of the capacity region, in addition to the sum-rate optimal
point, distributively. In slow varying environments, where the
multiple access users can be assumed to interact many times, the
repeated game formulation was shown to achieve all the boundary
points of the capacity region. Finally, we elucidated the
limitations of our game theoretic framework in vector multiple
access channels.

An interesting avenue for future work is to further investigate
the practical aspects of our framework. For example, a natural
extension is to consider the case with partial and/or distorted
channel state information by borrowing tools from game theory with
imperfect information.
\section{Acknowledgment}
The authors would like to thank Professor Wei Yu and Dr. Raul
Etkin for answering questions about their papers.

\end{document}